\documentclass[prb,twocolumn,amsmath,amssymb,floatfix]{revtex4}
\usepackage{graphicx}% Include figure files
\usepackage{bm}% bold math
\begin{document}
\draft

\newcommand{\be}{\begin{equation}}
\newcommand{\ee}{\end{equation}}
\newcommand{\beq}{\begin{eqnarray}}
\newcommand{\eeq}{\end{eqnarray}}

\title{Effect of quantum fluctuations on the dipolar motion of Bose-Einstein
condensates in optical lattices}

\author{Anatoli Polkovnikov and Daw-Wei Wang}

\address{Physics Department, Harvard University, Cambridge, MA 02138}
\date{\today}

\begin{abstract}
We revisit dipolar motion of condensate atoms in one-dimensional
optical lattices and harmonic magnetic traps including quantum
fluctuations within the truncated Wigner approximation. In the
strong tunneling limit we reproduce the meanfield results with a
sharp dynamical transition at the critical displacement. When the
tunneling is reduced, on the contrary, strong quantum fluctuations
lead to finite damping of condensate oscillations even at
infinitesimal displacement. We argue that there is a smooth
crossover between the chaotic classical transition at finite
displacement and the superfluid-to-insulator phase transition at
zero displacement. We further analyze the time dependence of the
density fluctuations and of the coherence of the condensate and
find several nontrivial dynamical effects, which can be observed
in the present experimental conditions.
\end{abstract}

%\pacs{PACS numbers:}

\maketitle
%%%%%%%%%%%%%%%%%%%%%%%%%%%%

The study of Bose-Einstein condensates of ultracold atoms has been
growing rapidly in recent years \cite{nature}. Loading bosenic
atoms into an optical lattice and enhancing the laser intensity,
it is possible to strongly suppress the kinetic energy of the
atoms resulting in a superfluid-to-Mott-insulator (SF-MI) quantum
phase transition~\cite{Jaksch,Bloch}: in the strong tunneling
limit the condensate is in the superfluid phase with finite phase
stiffness, while it is driven to the Mott-insulator phase and
loses the phase coherence because of quantum fluctuations if the
tunneling becomes small enough. Typical interference patterns
obtained in time-of-flight experiments cannot provide sufficient
information to see a sharp transition boundary~\cite{Bloch,roth}.

In one-dimensional systems the condensate dynamics has been
observed in a dipolar motion experiment~\cite{shaking_exp}, where
the center-of-mass (CM) of the condensate oscillates after a
sudden displacement of the magnetic trap with respect to the
optical lattice. This dynamics can be described by the meanfield
Gross-Pitaveskii equations (GPE) in the strong tunneling
regime~\cite{shaking_exp,stringari}. When the displacement is
beyond a certain critical value, the condensate oscillations are
overdamped~\cite{mi_exp} and the motion becomes completely
decoherent, indicating a classical localization
transition~\cite{niu, bishop}. Such a transition was recently
observed experimentally~\cite{inguscio}. It is plausible that as
the quantum fluctuations increase, there is a smooth crossover
between the quantum (SF-MI) and the classical transitions.
However, we emphasize that the first one is a second order
transition characterized by reversible phase coherence if the
system is driven to the insulating phase and then back to the
SF~\cite{Bloch}. On the other hand the classical transition is
irreversible~\cite{bishop} due to the chaotic excitations inside
the condensate cloud.

Motivated by this interesting and important question, in this
letter we investigate the quantum fluctuation effects on the CM
motion of a condensate in a parabolic potential in a
one-dimensional optical lattice using the truncated Wigner
approximation (TWA). In the strong tunneling (or weakly
interacting) regime, we reproduce the sharp dynamical transition
of a meanfield calculation~\cite{bishop} with a critical
displacement proportional to the square root of the tunneling
amplitude. When the tunneling is weak the quantum fluctuations
strongly modify the dynamics of the condensate in the following
ways: (i) the CM motion is damped even for an infinitesimal
initial displacement; (ii) the oscillations become overdamped at a
critical displacement, $D_c$, which is below the classical value
at the same tunneling; (iii) the condensate dipolar motion is
frictionless from one end to another with the period independent
of damping, while the superfluid fraction and the phase coherence
drop when the condensate passes through the center of the harmonic
well. (iv) We further show that at a given damping, the loss of
coherence is less for smaller displacements (or stronger quantum
fluctuations). So as we increase quantum fluctuations the
localization transition becomes more reversible. Our results hence
suggest that there is a smooth crossover between the classical
localization transition and the quantum superfluid-to-insulator
(SF-IN) transition~\cite{mott} as the displacement goes to zero.

The truncated Wigner approximation~\cite{Walls} has been well
known in quantum optics for a while. Recently, it was applied to
the systems of interacting bosons~\cite{TWA,ap1}. In
Ref.~[\onlinecite{ap1}] it was argued that TWA is equivalent to
the semiclassical approximation and naturally appears in the
quantum expansion of the time evolution of the system. The idea
behind TWA is that the expectation value of an observable $\Omega$
can be found according to:
\be
\langle\Omega(t)\rangle\approx \int d\psi_0d\psi_0^\star\,
P(\psi_0,\psi_0^\star)\Omega_{cl}(\psi(t),\psi^\star(t),t),
\label{1}
\ee
where $\psi$ and $\psi^\star$ denote bosonic fields obeying
classical discrete Gross-Pitaevskii equations of motion
\cite{GPeq,psg}:
\be
i{\partial \psi_j\over \partial t
}=-J(\psi_{j-1}+\psi_{j+1})+\frac{K}{2}j^2\psi_j+{U\over
2}|\psi_j|^2\psi_j
\label{2}
\ee
with the initial conditions $\psi(t_0)=\psi_0$ and
$\psi^\star(t_0)=\psi_0^\star$. The parameters $J$, $K$ and $U$ in
Eq.~(\ref{2}) denote the tunneling constant, the harmonic trap
curvature and the on-site interaction, respectively. The function
$P(\psi_0,\psi_0^\star)$ is a Wigner transform of the initial
density matrix, and can be interpreted as the probability of
having particular initial conditions~\cite{note1}. Finally
$\Omega_{cl}$ is the Weyl symbol of the operator $\Omega$
evaluated on the classical fields $\psi(t)$ and $\psi^\star(t)$.
It is important to realize that TWA considerably improves
Bogoluibov's theory, especially if the classical dynamics becomes
unstable~\cite{ap1}. The way we implement TWA in this letter is
outlined in Ref.~[\onlinecite{ap1}].

\begin{figure}[ht]
\includegraphics[width=7.5cm]{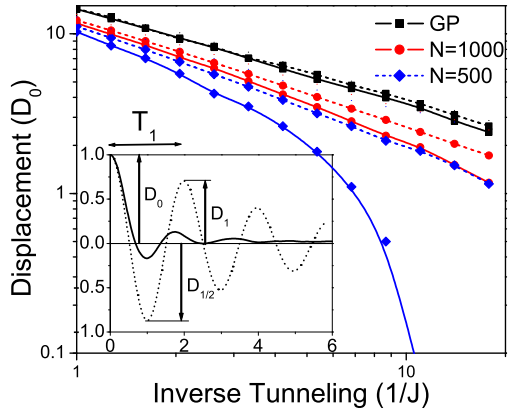}
\caption{Dynamical phase diagram for $UN=50$, and $K=0.02$ (the
occupancy of the central site, $N_0$, is approximately $5\%$ of
the total number of bosons, $N$). The solid and the dashed lines
correspond to damping $\gamma=0.11$ and $\gamma=0.36$
respectively. The large separation between the two curves for
stronger quantum fluctuations (smaller $N$) implies broadening of
the transition. The inset shows the typical temporal CM
oscillations for $\gamma\approx 0.36$ (dashed line) and
$\gamma\approx 2.1$ (solid line). }
\label{phase_diag}
\end{figure}
In a dipolar motion experiment, the condensate is initially
($t<0$) prepared in the superfluid ground state. At $t=0$ the trap
position is suddenly displaced by the distance $D_0$ from the
origin and the condensate starts to move. To find the Wigner
transform $P(\psi_0,\psi_0^\ast)$ of the interacting ground state,
we start from the noninteracting Hamiltonian ($U=0$), where the
function $P(\psi_0,\psi_0^\ast)$ can be trivially computed, and
then adiabatically increase $U$ to the actual value~\cite{ap1}. We
check that the increase of the interaction is slow enough so that
the final result is not affected by this procedure. The further
implementation of Eq.~(\ref{1}) is based on the Monte-Carlo
averaging of the $\Omega_{cl}(\psi(t),\psi^\star(t),t)$ over the
different initial conditions weighted by the probability
$P(\psi_0,\psi_0^\ast)$.

In Fig.~\ref{phase_diag} we show the calculated dynamical phase
diagram in terms of the inverse tunneling $1/J$ and the
displacement $D_0$. The contours correspond to the constant
damping of the CM oscillations, $\gamma\equiv \ln (D_0/D_1)$,
where $D_0$ and $D_1$ are the CM positions at $t=0$ and $t=T_1$
(see the inset). The solid and the dashed lines correspond to the
damping of $\gamma=0.11$ and $0.36$ respectively
\cite{note_frequencydamping}. The top two curves correspond to the
Gross-Pitaevskii result ($D_c\propto\sqrt{J}$), where the
transition is so sharp that they almost coincide. When we increase
quantum fluctuations reducing the total number of bosons $N$ (but
keeping the product $UN$ the same, which is equivalent to fixing
the chemical potential~\cite{chemicalpotential}), we find: (i) for
a given $J$, the damping of the oscillations occurs at a smaller
displacement than in the classical case, and the transition
becomes broader for smaller $N$. (ii) If the tunneling is small
enough (say $J^{-1}>10$ for $N=500$), then even an infinitesimal
displacement results in considerable damping of the CM motion.
Although because of the time-consuming computation we are able to
trace only $\gamma=0.11$ contour for $N=500$ to zero, it is clear
that such a behavior should be generally true for any given
$\gamma$, if $J$ becomes small enough. For the parameters chosen
in Fig.~\ref{phase_diag}, the SF-IN transition \cite{mott} occurs
at $J^{-1}_c\sim N_0/U\sim 10^3$. This is much larger than the
range of tunneling where we perform the calculations so that the
results we obtained in this Letter within TWA are expected to be
fairly reliable. We anticipate that contours of larger damping
will terminate much closer to the SF-IN transition.
\begin{figure}[ht]
\includegraphics[width=7.5cm]{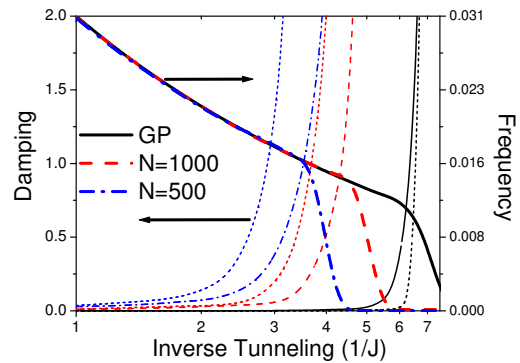}
\caption{Damping (thin lines) and the oscillation frequency (thick
lines) v.s. inverse tunneling for $D_0=5$. Dotted lines show
damping for a twice as large system with the displacement also
scaled by a factor of two (see the text). }
\label{damping}
\end{figure}
In Fig.~\ref{damping} we show the damping of the CM motion as a
function of the inverse tunneling at $D_0=5$. Due to the finite
size effects, GP solution also has some finite damping close to
the transition point ($\gamma=\infty$). Clearly the damping
becomes broader and the transition point shifts to the lower
tunneling when quantum fluctuations are enhanced. To affirm the
quantum effects in the thermodynamic limit, we also show the
results (dotted lines) for a twice as large system ($K\to K/4$ and
$N\to 2N$, keeping the same chemical
potential~\cite{chemicalpotential}). The displacement was also
scaled by a factor of two in order to fix the same condensate
maximum velocity. We find that the damping of the GP solution
becomes a sharper function of the tunneling (but diverges at the
same tunneling amplitude) while the quantum CM motion is destroyed
at even larger tunneling! This result reflects the fact that the
GP solution has only one length scale associated with the
condensate size. In the quantum case there appears another scale,
associated with generation of quasiparticles out of the
condensate. Therefore, the finite damping of the CM motion shown
in Figs.~\ref{phase_diag} and~\ref{damping} exists even in the
thermodynamic limit if the condensate velocity (proportional to
$D\sqrt{K}$) is kept finite. In Fig.~\ref{damping}, we also show
the dependence of the oscillation
frequency~\cite{note_frequencydamping}($\omega$) on tunneling.
Surprisingly we find that $\omega$ does not deviate from the GP
result even for the relatively strong damping as long as the CM
motion remains underdamped.

\begin{figure}[ht]
\includegraphics[width=7.5cm]{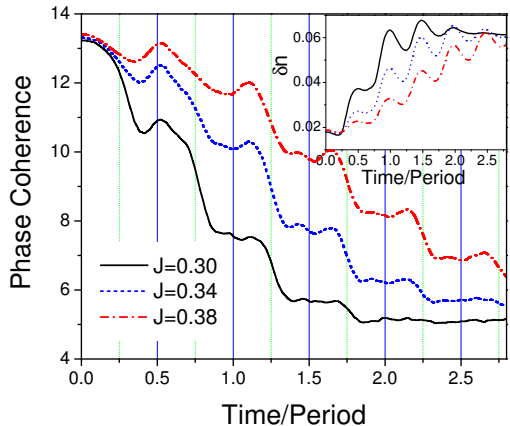}
\caption{Phase coherence (main) and density fluctuations (insert)
versus time for the same displacement ($D_0=5$) and different
tunneling constants. Note that both quantities
have been rescaled to be the same at $t=0$ for the convenience of comparison.}
\label{density_coherence}
\end{figure}

In Fig.~\ref{density_coherence} we show the standard deviation of
the boson density, $\delta n\equiv N^{-1} \sqrt{\sum_j\left( \langle
N_j^2\rangle-\langle N_j\rangle^2\right) }$, and the phase
coherence, $C=N^{-1}\,{\rm Max}_q\left[\sum_{jl}\langle
\psi_j^\dagger \psi_l\rangle ,e^{i(j-l)q}\right]$, as a function
of time for $N=1000$, $D_0=5$ and different tunneling amplitudes.
We first note that both $\delta n$ and $C$ only slightly depend on
time if the condensate is near the turning points (i.e. at
time/period=$0.5,1,1.5,\dots$), while they change significantly
when the condensate has its maximum velocity at the center of the
parabolic well (time/period=$0.25,0.75,1.25,\dots$). The sharp
increase of the density fluctuations (and the sharp loss of the
coherence) can be interpreted as fast generation of the incoherent
quasi-particles
with a simultaneous decrease of the superfluid fraction. Combining
this with the fact that the oscillation frequency does not deviate
from its classical (GP) value even for the strong damping, we
argue that the whole CM motion can be interpreted within Landau's
two-fluid model \cite{landau}: the superfluid component oscillates
frictionlessly and is well-described by GP equations, while the
normal fluid component has a strong damping and becomes localized
near the center of the parabolic well. The former ensures the same
oscillation frequency as the classical (GP) value, while the
latter causes the damping of the amplitude of the CM motion. The
generation of quasi-particles continues in each cycle of the
oscillations until the coherent motion stops
(Fig.~\ref{density_coherence}). If the initial superfluid fraction
is too small due to the large quantum depletion, the
quasi-particle generation can be so efficient that the CM motion
becomes overdamped ($\gamma=\infty$).

\begin{figure}[ht]
\includegraphics[width=7.5cm]{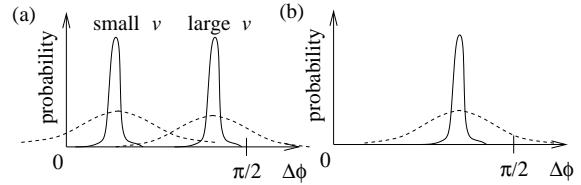}
\caption{Schematic figures of the distribution of the phase
gradient for the classical (solid) and quantum (dashed) initial
states for (a) the same tunneling and different displacements, and
for (b) the same displacement and different tunneling. The sharp
peak distribution for the classical case indicates that the local
velocity of the condensate does not fluctuate; $\pi/2$ is the
critical phase gradient for the classical transition. }
\label{phase_distribution}
\end{figure}

We can qualitatively understand quantum effects on the dynamical
transition using a schematic plot of the probability distribution
of the phase difference between two nearest sites as shown in
Fig.~\ref{phase_distribution}.
Given the same tunneling amplitude [Fig.
\ref{phase_distribution}(a)], the classical distribution of phase
gradients peaks at the CM momentum, $v$. In the meanfield picture,
the CM motion of the condensate is frictionless~\cite{bishop} if
$v$ is smaller than the critical value, $v_c=\pi/2$, while it is
completely destroyed if $v>v_c$. If we include quantum
fluctuations, then the local phase gradient distribution becomes
broader, and therefore there is a finite probability to have a
large phase difference in a given link even if $v<v_c$, which
results in the finite damping of the CM motion. Similarly, at a
given $v$ [Fig.~\ref{phase_distribution}(b)] smaller values of $J$
lead to a broader distribution of the phase gradient, and hence to
larger damping.
\begin{figure}[ht]
\includegraphics[width=7.5cm]{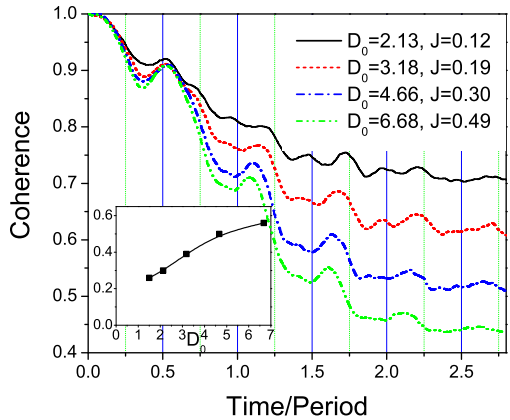}
\caption{Phase coherence (renormalized to the same value
at $t=0$) versus time for different initial displacement $D_0$
(or the associate tunneling $J$ as indicated)
with the same damping
$\gamma=0.36$ of $N=1000$ (i.e. along the dashed line with circles of
Fig. \ref{phase_diag}). The inset shows the loss of
coherence ($(C(t=0)-C(t=\infty))/C(t=0)$) of these data as a function of
the displacement. Clearly smaller $D_0$ (and
hence $J$) results in a smaller loss of coherence of the
condensate.}
\label{same_damping}
\end{figure}

Now let us look closer to the connection between the classical
transition at finite displacement and the quantum SF-IN
transition~\cite{mott} at zero displacement. In
Fig.~\ref{same_damping}, we show the phase coherence as a function
of time for different initial displacements but at the same
damping, $\gamma=0.36$. We note that the coherence saturates at a
{\it finite} value, which is closer to the initial one for
stronger quantum fluctuations (smaller $J$ and $D_0$) as shown in
the inset. These results suggest that along the $\gamma=\infty$
contour the coherence loss should also approach zero if the
displacement gets smaller~\cite{note2}, indicating a smooth
crossover between the dynamical irreversible transition at finite
displacement and the reversible quantum phase transition at zero
displacement. We note that such a dipolar motion can be also used
to investigate the superfluid to Bose glass insulating phase when
a bound random potential is applied~\cite{disorder}.

Our analysis also implies that in uniform systems quantum
fluctuations lead to the damping of the condensate current state
even if the velocity is below the GP critical value~\cite{bishop}.
A similar effect was predicted for the current decay in quasi
one-dimensional superconductors due to thermal
fluctuations~\cite{LA}, and more recently for the atomic system
with a moving defect~\cite{buchler}. We will investigate
mechanisms of the current decay in uniform systems in separate
publications. The results predicted in this letter can be directly
tested  experimentally. For example, it should be possible to
observe the damping of the condensate motion at small
displacements (Fig. \ref{damping}) and the ladder-like structure
in the coherence or in the number variance (Fig.
\ref{density_coherence}) as a function of time.

In summary, we studied the effects of quantum fluctuations on the
dynamical properties of atomic condensates in 1D optical lattices
with a parabolic confinement potential. We showed that the quantum
fluctuations are very important even {\it far from} the
superfluid-to-insulator transition boundary. We further
demonstrate that the dynamical localization transition, which has
a purely classical origin, can be smoothly connected with the
static quantum SF-MI phase transition. Our results give a number
of predictions on the condensate dynamics which can be directly
tested in experiments.

We acknowledge useful discussions with E.~Altman, E.~Demler,
B.~Halperin, M.~Lukin, and P.~Zoller. This work was supported by
US NSF grants DMR-0233773, DMR-0231631, and DMR-0213805.

%%%%%%%%%%%%%%%%%%%%%%%%%%%%%%%%%%%%%%%%%%%%%%%%

%%%%%%%%%%%%%%%%%%%%%%%%%%%%%%%%%%%%%%%%

%%%%%%%%%%%%%%%%%%%%%%%%%%%%%%%

\begin{thebibliography} {}
\bibitem{mott}
We note that due to the inhomogeneity of the system, it is hard to
define a Mott insulator and its transition to the superfluid.
Therefore we suggest that the overdamped regime at zero
displacement should be associated with the insulator regime in a
more general sense.
\bibitem{note1}
This interpretation is not precise, since $P$ is not positively
defined. For details, see Ref. [\onlinecite{ap1}].
\bibitem{chemicalpotential}
Note that in this weak tunneling regime, the condensate size is
not sensitive to the tunneling (so called Thomas Fermi limit). The
density profile can be estimated by $N_j=N_0(1-K j^2/\mu)$ where
$N_0$ is the number of atoms in the central well and the chemical
potential $\mu=UN_0$ is determined by the total number of bosons
$N$.
\bibitem{note2}

We note that the coherence loss during the dynamical process is
best defined by the value after adiabatically tuning the
interaction $U$ to zero. But the results we present here are clear
enough to show its qualitative behavior.
\end{thebibliography}
\end{document}